\documentstyle[aps,twocolumn,epsf]{revtex}
\title{Semiclassical spectra from periodic-orbit clusters in a
mixed phase space}
\author{Henning Schomerus and Fritz Haake}
\address{Fachbereich Physik,
	Universit\"at-Gesamthochschule Essen,
		D-45117 Essen, Germany}

\begin{document}
\draft
\maketitle
\begin{abstract}
We determine semiclassical quasienergy
spectra from periodic orbits for a system with a mixed phase space, the
kicked top. Throughout the transition from integrability to well
developed chaos the standard error incurred for the quasienergies is a
small percentage of their mean spacing. Such fine accuracy
does not even require the angular momentum quantum number $j\propto
1/\hbar$ to be large; it already prevails for $j=1,2,3$.
To open the way towards a reliable spectrum the conventional trace
formula \`a la Gutzwiller has to be extended by including (i) ghosts,
i.e., complex predecessors of bifurcating orbits and the possibility of
Stokes transitions undergone by such complex orbits and (ii)
collective contributions of clusters of periodic orbits near
bifurcations. Even bifurcations of codimension higher than one must be
reckoned with by accounting for the clusters involved through the
appropriate diffraction integrals.
\end{abstract}


Since Gutzwiller's seminal work \cite{Gutzwiller:1971,Gutzwiller:1990}
it is known that
the level density of autonomous hyperbolic systems
can be semiclassically approximated by a sum
of individual contributions from periodic orbits.
Similarly, the spectrum of integrable systems
may be calculated semiclassically by EBK quantization.
Gutzwiller's result was later extended
to periodically driven systems \cite{Tabor:1983,Junker:Leschke:1992}
whose stroboscopic period-to-period evolution is generated by a
unitary Floquet operator $F$ with unimodular eigenvalues
$e^{-i\varphi_i}$. The quasienergies $\varphi_i$ are encoded
in the traces $\mbox{tr}\,F^n$, $n=1,2,\ldots$,
which are approximated as
\begin{equation}\label{trFindiv}
\mbox{tr}\,F^n=\sum_{\mbox{\scriptsize p.o.}}^{\mbox{\scriptsize period }n}
\frac{n_0}{|2-\mbox{tr}\,M|^{1/2}}
\exp\left(i\frac S \hbar-i\frac \pi 2\nu\right)
\end{equation}
for systems with a single classical degree of freedom.
Each period-$n$ orbit provides a summand determined by its primitive
period $n_0$, the action $S$, the Maslov index $\nu$, and the trace of
the linearized map $M$.

The Gutzwiller type result (\ref{trFindiv}) can be derived
from the integral representation
\begin{equation}\label{trFint}
\mbox{tr}\,F^n=\int\frac{dq^\prime\,dp}{2\pi\hbar}
\left|S_{q^\prime p}\right|^{1/2}
e^{\frac i\hbar [S(q^\prime,p;n)-q^\prime p]-i\frac \pi 2 \mu}\;,
\end{equation}
which involves, besides the Morse index $\mu$, the action $S(q^\prime,p;n)$;
the latter generates the  $n$-step map
$(q,p)\to(q^\prime,p^\prime)$ through $S_p=q$, $S_{q^\prime}=p^\prime$,
where the indices on $S$ denote partial derivatives.
The stationary-phase approximation leading from the integral (\ref{trFint})
to the periodic-orbit sum (\ref{trFindiv}) is a
sensible one provided all stationary points are well separated.

Generic systems, however, are neither integrable nor chaotic but come
with a mixed phase space with stability islands
residing in a sea of chaotic motion. The present Letter is devoted to the
semiclassical evaluation of the quasienergy spectrum in that situation.
In particular, we shall be concerned with the transition from regular to
predominantly chaotic behavior as a suitable control parameter is
varied. Such a transition involves complex changes of the
stability islands when periodic orbits arise, disappear, or coalesce at
bifurcations. The bifurcations generically encountered upon varying a
single parameter and therefore said to have codimension one have been
classified by Meyer \cite{Meyer:1970}. The simplest type is the
so-called tangent bifurcation at which a pair of new classical periodic
orbits is born (or coalesces and disappears for the opposite sense of
change of the control parameter). The general cases are
period-$m$ bifurcations where a ``central'' orbit of period $l$
coalesces with satellites
of $m$-fold period $n=m l$.
At the bifurcation the $n$-th iterate of the
linearized map $M$ is the identity close to the coalescing orbits
and gives, for one degree of freedom,
the condition $\mbox{tr}\,M^n=2$. Clearly, a tangent bifurcation
may be seen as the special case $m=1$. 

As a dynamical system is driven
through a sequence of bifurcations towards globally chaotic behavior the
simple Gutzwiller type trace formula (\ref{trFindiv}) ceases to be a
reasonable approximation to the integral (\ref{trFint}), mostly since
different periodic orbits approach one another so closely as to become
incapable to yield independent additive stationary-phase contributions
to the integral (\ref{trFint}).
Right at a bifurcation individual contributions to 
$\mbox{tr}\,F^n$ even diverge.
To construct a ``collective''
contribution \cite{Ozorio:Hannay:1987,Ozorio:1988}
in the neighborhood of a bifurcation one must approximate
the action function $S$ in (\ref{trFint}) by a suitable normal form
whose stationary points yield the cluster of classical periodic points
related to the bifurcation; the ensuing ``diffraction catastrophe
integral'' then constitutes a cluster contribution to the trace
$\mbox{tr}\,F^n$ in question. For some recent progress with several
diffraction integrals relevant for our present study we refer the reader
to \cite {Sieber:1996a,Schomerus:Sieber:1997a,Schomerus:1997a}.

When periodic orbits disappear as a control parameter passes
through a critical value the nonlinear classical map loses a number of real
solutions in favor of complex ones. A complex ``ghost'' orbit has
no classical significance but does yield a saddle-point contribution
\cite{Kus:Haake:1993b} to the integral (\ref{trFint}). In the immediate
neighborhood of the said bifurcation the ghosts again form
(part of) a cluster which must be treated by an appropriate diffraction
integral; then one obtains a uniform interpolation between the asymptotic
behaviors on both sides of the bifurcation, due to the saddles for the
ghosts on one side and the stationary phases for the real successors on
the other side. The simplest and best known such case arises for a tangent
bifurcation for which the diffraction integral takes the familiar Airy
function form \cite{Ozorio:Hannay:1987,Ozorio:1988,Kus:Haake:1993b}.

A ghost orbit often makes itself felt surprisingly far away from the
bifurcation from which it originates, since the imaginary part of its
action [which in principle entails exponential suppression through the
factor $\exp{(-{\rm Im}\,S/\hbar})$] may decay slowly as one steers the
dynamics away from the bifurcation. We in fact find that a ghost
frequently loses its weight through another mechanism, i.e., the
so-called Stokes transition after which the corresponding saddle of
the integrand in (\ref{trFint}) can no longer be reached by deforming
the original contour of integration to one of steepest descent without
crossing a singularity. The transition is encountered when the real
parts of the action are identical for a ghost ($-$) and another ``dominant''
orbit in
its vicinity ($+$), with ${\rm Im}\, S_+<{\rm Im}\, S_-$. The phenomenon
has been investigated in \cite{Berry:1989,Boasman:Keating:1995}, where a
uniform approximation of the suppression factor is given.
Incidentally, the Stokes phenomenon also implies that only ghosts with
${\rm Im}\,S>0$ are relevant.

Another surprise will be incurred below, in our search for reliable
semiclassical approximations to the traces $\mbox{tr}\,F^n$:
Classically nongeneric bifurcations become quantitatively important.
These have codimension two, i.e., could be located only by controlling
two parameters. Even though we shall be concerned with varying but a
single parameter and never actually hit such a codimension-two
bifurcation it seems typical to get sufficiently close for collective
treatments of all participating orbits to become necessary.

Leaving generalities for the moment we now turn to pursuing our goal for a
periodically kicked angular momentum often referred to as kicked top
\cite{Bibel,Haake:Kus:1987,Kus:Haake:1993a,Braun:Gerwinski:1996}.
The angular momentum $\bf J$ involved has
components obeying the usual commutation rules $[J_x,J_y]=iJ_z$,
etc. The squared angular momentum
${\bf J}^2=j(j+1)$ is conserved and with the quantum number
$j=\frac{1}{2},1,\frac{3}{2},2,\ldots$ fixed the
Hilbert space is assigned the dimension $2j+1$. That dimension also
plays the role of the inverse of Planck's
constant such that classical behavior is attained in the limit $j\to
\infty$. We shall work here with the particular top whose dynamics is
a sequence of rotations by angles
$p_i$ alternating with torsions
of strength $k_i$ as described by the Floquet operator
\begin{eqnarray}
F&=&\exp(-i k_z\frac {J_z^2}{2j+1}-ip_z J_z)
\exp(-i p_y J_y) \nonumber
\\
&&\times
\exp(-i k_x\frac {J_x^2}{2j+1}-ip_x J_x)\;.
\end{eqnarray}
The corresponding classical map may be obtained by writing out the
stroboscopic dynamics of the rescaled angular momentum vector
${\bf X}={\bf J}/(j+ 1/2)$
in the Heisenberg picture and then, with the limit
$j\to \infty$ in mind, degrading $\bf X$ to a c-number vector.
The classical phase space is thus revealed as the unit sphere ${\bf X}^2=1$.
The classical stroboscopic map is easily written as
the sequence of three rotations, one about the $x$-axis by the angle
$p_x+k_xx$, the second by $p_y$ about the $y$-axis, and the last one by
$p_z+k_zz$ about the $z$-axis.

\begin{figure}[t]
\epsfbox{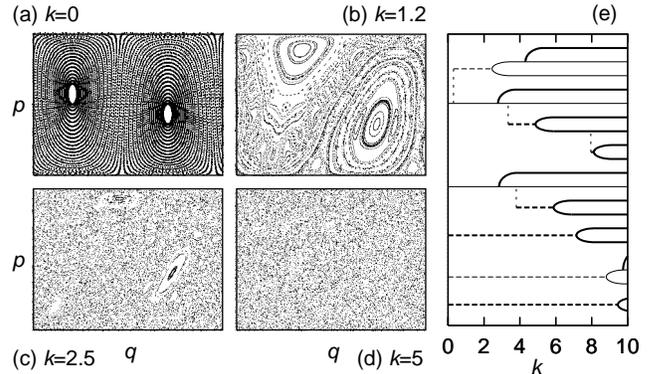}
\caption{(a)-(d): Phase space portraits for the kicked top.
The spherical phase space is parameterized by the azimuthal
angle $\varphi\equiv q$ and the Cartesian coordinate $z\equiv p$.
Varying
the control parameter $k$ from $0$ to $5$,
the system undergoes the transition from
integrability through mixed phase space to well developed chaos.
(e):
Bifurcation tree including periodic orbits
of period one (thin lines) and two (thick lines). Solid lines are real orbits,
broken lines
ghost solutions with $\mbox{Im}\,S>0$.
Ghost lines end at Stokes transitions, where
vertical lines connect them to the dominant orbit.
This is not indicated for strongly suppressed ghosts with large
${\rm Im}\,S$.}
\label{fig1}
\end{figure}

We perform a one-parameter study of the top, holding the $p_i$ fixed
($p_x=0.3$, $p_y=1.0$, $p_z=0.8$) while varying the control parameter
$k\equiv k_z=10k_x$ in the range $0\leq
k\mathop{\raisebox{-0.8ex}{\shortstack{$<$\\[-0.4ex]$\sim$}}}10$. For
$k=0$ we incur a pure linear rotation, and thus classical integrability. Only
two primitive periodic orbits then arise, i.e., fixed points located at
the intersections of the rotation axis with the unit sphere. For $k=5$
the phase-space portrait in Fig.\ \ref{fig1} displays well developed
chaos.

Bifurcations are found numerically by solving the equation
${\rm tr}\,M^m=2$ simultaneously with $p=p^\prime$, $q=q^\prime$
for the triple $(q, p,
k)$. All periodic orbits existing for a certain value of $k$ can be
picked up by going through the sequence of bifurcations as $k$ is swept
up from zero to its current value. Figure \ref{fig1}(e) displays the
bifurcation tree thus obtained, showing twenty orbits of length one and
two subsequently used in the evaluation of $\mbox{tr}\,F^2$ for $k<10$.

\begin{figure}[t]
\epsfbox{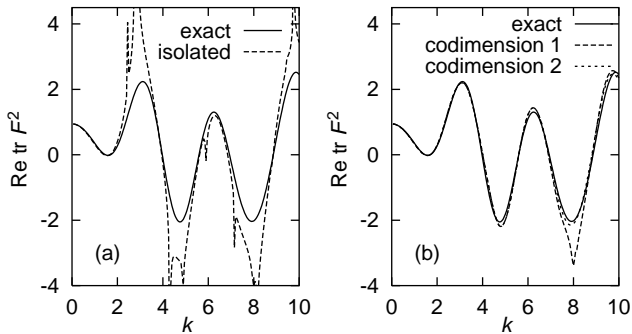}
\caption{The real part of the quantum-mechanically exact
$\mbox{tr}\,F^2(k)$ is compared
with various levels of semiclassical approximations. In (a),
individual contributions 
from all real orbits
and ghosts are summed up and
Stokes transitions taken
into account. In (b), collective contributions of orbit clusters are
used to regularize the behavior close to bifurcations. Clusters 
connected to codimension-one bifurcations are found to be insufficient at
$k\approx 8$. Enlarging the clusters further gives an
accurate approximation.}
\label{fig2}
\end{figure}

The divergence of individual contributions at bifurcations
is illustrated in Fig.\ \ref{fig2}(a),
which displays the quantum-mechanically exact $\mbox{tr}\,F^2(k)$
for $j=3$ together with the sum of individual contributions from real
periodic orbits and ghosts, eq.\ (\ref{trFindiv}).
Stokes transitions are
taken into account for those ghosts that are not sufficiently suppressed
by having a large imaginary part of the action.

We proceed further to include collective contributions that regularize
the behavior close to bifurcations. The broken line in Fig.\ \ref{fig2}(b)
is our result when one groups the orbits according to the
codimension-one bifurcations in which they participate (tangent
bifurcations of orbits with primitive period one and two and
period-doubling bifurcations of orbits with period one) and summing up
the corresponding contributions, using the closed formulae from
\cite{Schomerus:Sieber:1997a}.

A typical codimension-one cluster is that of the two period-one orbits
that come into existence via the tangent bifurcation at $k=2.44$. One of
the orbits is unstable while the other, initially stable, becomes
unstable in a period-doubling bifurcation at $k=4.30$; as it does, a
stable period-two orbit shows up as a satellite. Close to this
bifurcation one has thus another cluster, formed by the satellite and
the period-one orbit that changes its stability. Somewhere in between
these two bifurcations one clearly has to rearrange the clusters.
Unfortunately, the regrouping allows for ambiguities when an orbit is
involved in several subsequent bifurcations. Whenever a regrouping is
found necessary we choose its location along $k$ so as to minimize the
discontinuity in the approximated trace, taken as a function of $k$. In
most cases, as in the example under discussion,
the remaining discontinuity is tiny. In
three situations, however, we have to avoid such patchwork and to enlarge
our clusters to include orbits that are involved in subsequent
bifurcations. In two cases the clusters stem from bifurcations of
codimension two, where one would have to control two parameters to let
all participating orbits coalesce.

In one of the codimension-two situations, endangering the semiclassical
approximation of
$\mbox{tr}\,F^2$ around $k\approx 8$, a third orbit of equal length is
found in close neighborhood to a pair of orbits that participate in a
tangent bifurcation at $k=8.12$, and the Stokes transition rendering the
ghost orbits irrelevant occurs at $k=7.98$.
This type of clusters
formed by three orbits of equal period is also frequently encountered
for $\mbox{tr}\,F^3$. It can be described by the normal form
\begin{equation}
S^{(1)}(q^\prime,p)=q^\prime p-\varepsilon q^\prime
-a{q^\prime}^3-b{q^\prime}^4-\frac{\sigma}2p^2\;,
\end{equation}
with $\sigma=\pm 1$. This describes three orbits, two of which bifurcate at
$\varepsilon=0$. A uniform approximation is obtained by introducing
$S^{(1)}$ into the exponent of eq.\ (\ref{trFint}) and expanding
$|S_{q^\prime p}|^{1/2}=1+A q^\prime+B{q^\prime}^2$. Here $A$ and $B$
are determined by requiring that the resulting contribution
\begin{eqnarray}
C^{(\rm cluster)}&=&\frac{1}{\sqrt{2\pi\hbar}} \int
dx\,(1+Ax+Bx^2) \\&& \times \exp\left(\frac{i}{\hbar}(-\varepsilon x-a
x^3-bx^4)-i\frac{\pi}{2}(\mu+\frac \sigma 2)\right) \nonumber
\end{eqnarray}
has the right stationary-phase limit as $\hbar \to 0$,
which gives three individual contributions of the type encountered in eq.\
(\ref{trFindiv}). The integral can be expressed by Pearcey's function
and its derivatives \cite{Pearcey:1946}. It turns out that this
expression also correctly treats the Stokes transition of the complex
saddles.

Upon treating the codimension-two event as discussed we obtain the
dotted line in Fig.\ \ref{fig2}(b) for the second trace,
$\mbox{tr}\,F^2$. This ultimate level of approximation reproduces the
exact result quite nicely.

The other codimension-two case encountered is that of a
tripling bifurcation close to a tangent bifurcation of the satellite
period-three orbit. This involves another satellite of period three that
can be taken into account by an extended normal form, as is discussed in
greater detail in \cite{Schomerus:1997a}, where a uniform
approximation is given. It becomes relevant in the evaluation of
$\mbox{tr}\,F^3$, as does also the third scenario of higher codimension
where
a tangent bifurcation
of period-three orbits takes place 
on a broken torus formed by another pair of
period-three orbits. The result for $\mbox{tr}\,F^3$ is considerably
improved by treating all four orbits collectively, using the
contribution
\begin{eqnarray*}
&&C^{(\rm cluster)}=\int_0^{2\pi}\!\!\!
\frac{d\varphi}{\sqrt{2\pi\hbar}}
(A+B\cos(\varphi+\varphi_0)+C\cos 2\varphi) \nonumber\\
&&\quad\times 
\exp\left(\frac{i}{\hbar}(a\cos(\varphi+\varphi_1)+b\cos 2\varphi)
-i\frac{\pi}{2}(\mu+\frac \sigma 2)\right)\;,
\end{eqnarray*}
where all coefficients are determined to yield the correct
stationary-phase limit.

With help of these  collective contributions
the traces $\mbox{tr}\,F$ and
$\mbox{tr}\,F^3$ come out with a quality comparable to that of
$\mbox{tr}\,F^2$.

In general, the Floquet
operator $F$ acts as an $N\times N$ matrix with $N=2j+1$ whose eigenvalues
$e^{-i\varphi_i}$ are determined by the set of traces with, for integral
$j$, $n=1\ldots j$.
For $j=3$ the first three traces thus provide sufficient information to
retrieve all seven quasienergies.
Indeed, from these traces one obtains the first half of
the coefficients in the secular polynomial $\mbox{det}(F-z)=\sum_{n=0}^N
a_{N-n}(-z)^n=0$ using Newton's formulae \cite{Newton}; the second half
follows from the unitarity of $F$ which entails the so-called
self-inversiveness \cite{Bogomolny:Bohigas:1992,HKSSZ},
$a_{N-n}=a_n^*a_N$.
We benefit from the fact that
$a_N=\mbox{det}\,F$, needed to exploit the
self-inversiveness of the polynomial, is accessible semiclassically: For
the top,
$\mbox{det}\,F$ factorizes
into a product of determinants of pure rotations and torsions, and each
integrable factor can be treated individually by EBK quantization,
which even gives
the exact result, $\mbox{det}\,F=\exp[-\frac i3 j(j+1)(k_x+k_z)]$.

\begin{figure}[t]
\epsfbox{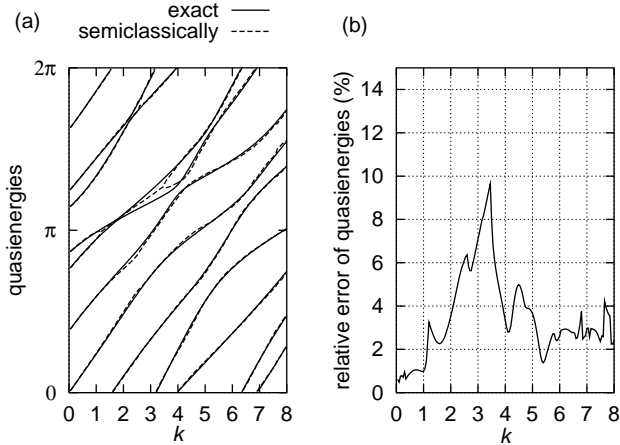}
\caption{(a) Exact and semiclassical level curves as a function of $k$.
(b) Relative error $\frac{\Delta \varphi}{2\pi/(2j+1)}$ of the
quasienergies.}
\label{fig3}
\end{figure}

The quasienergies $\varphi_i(k)$ from the semiclassically approximated
secular polynomial are plotted together with the exact ones 
in Fig.\ \ref{fig3}(a). Noteworthy is
the coalescence of two semiclassical phases mimicking  a close
avoided crossing of exact ones. The corresponding eigenvalues
cease to be unimodular there.
Self-inverse polynomials are capable of such behavior; to impose
unitarity on them certain additional restrictions have to be fulfilled
by the set of traces which are not automatically
respected by the semiclassically approximated ones.
As a quantitative measure of accuracy
we employ the standard deviation between exact and semiclassical
quasienergies,
\begin{equation}
\Delta \varphi =
\sqrt{\frac1{2j+1}\sum_{i=1}^{2j+1} \left(\varphi^{(\rm
sc)}_i-\varphi^{(\rm qm)}_i\right)^2} \;,
\end{equation}
which is shown in Fig.\ \ref{fig3}(b). The error is a small single-digit
percentage of the
mean spacing $2\pi/(2j+1)$, a little higher only in the short $k$ interval
where the two semiclassical phases mentioned are degenerate.

Similarly small is the error we have incurred for $j=2$ and even $j=1$.
Semiclassical behavior of the spectrum begins to prevail for
surprisingly small values of $j$ indeed, fortunately so before
with increasing $j$ the infamous exponential proliferation of
periodic orbits would render semiclassical work cumbersome.

Our accuracy is comparable to the one previously found with
different semiclassical strategies that avoid periodic orbits
\cite{Braun:Gerwinski:1996,Gerwinski:Haake:1995}.

To summarize, we have demonstrated that the spectrum of
the kicked top, which has a mixed phase space, can be calculated
semiclassically from periodic classical orbits, provided one improves in
several ways upon the trace formula for hyperbolic systems, eq.\
(\ref{trFindiv}). The modified trace formula complements the
contributions from isolated real periodic orbits by accounting for
isolated ghosts as well as clusters of orbits associated with
bifurcations of codimension one, two,\ldots~.

The authors have the great pleasure to thank P.\ Braun,
J.\ P.\ Keating, M.\ Ku\'s, M.\ Sieber,
U.\ Smilanski, S.\ Tomsovic, and  G.\ Wunner for enlightening discussions.
Support by the Sonderforschungsbereich
`Unordnung und gro{\ss}e Fluktuationen' of the Deutsche Forschungsgemeinschaft
is gratefully acknowledged.


\begin{references}

\bibitem{Gutzwiller:1971}
M.~C. Gutzwiller, J. Math. Phys. {\bf 12},  343  (1971).

\bibitem{Gutzwiller:1990}
M.~C. {Gutzwiller}, {\em Chaos in Classical and Quantum Mechanics} (Springer,
  New York, 1990).

\bibitem{Tabor:1983}
M. Tabor, Physica D {\bf 6},  195  (1983).

\bibitem{Junker:Leschke:1992}
G. Junker and H. Leschke, Physica D {\bf 56},  135  (1992).

\bibitem{Meyer:1970}
K.~R. Meyer, Trans. Am. Math. Soc. {\bf 149},  95  (1970).

\bibitem{Ozorio:Hannay:1987}
A.~M. {Ozorio de Almeida} and J.~H. {Hannay}, J. Phys. A {\bf
  20},  5873  (1987).

\bibitem{Ozorio:1988}
A.~M. {Ozorio de Almeida}, {\em {H}amiltonian Systems: Chaos and Quantization}
  (Cambridge University Press, Cambridge, 1988).

\bibitem{Sieber:1996a}
M. {Sieber}, J. Phys. A {\bf 29},  4715  (1996).

\bibitem{Schomerus:Sieber:1997a}
H. Schomerus and M. Sieber, submitted to J. Phys. A,
preprint chao-dyn/9701022 (to be published).

\bibitem{Schomerus:1997a}
H. Schomerus, preprint chao-dyn/9703003 (to be published).

\bibitem{Kus:Haake:1993b}
M. Ku{\'s}, F. Haake, and D. Delande, Phys. Rev. Lett. {\bf 71},  2167  (1993).

\bibitem{Berry:1989}
M.~V. Berry, Proc. R. Soc. Lond. A {\bf 422},  7  (1989).

\bibitem{Boasman:Keating:1995}

P.~A. Boasman and J.~P. Keating, Proc. R. Soc. Lond. A {\bf 449},  629  (1995).

\bibitem{Bibel} F.~Haake, {\sl Quantum Signatures of Chaos}, Springer,
Berlin, 1990.

\bibitem{Haake:Kus:1987}
F. Haake, M. Ku{\'s}, and R. Scharf, Z. Phys. B {\bf 65},  381  (1987).

\bibitem{Kus:Haake:1993a}
M. Ku{\'s}, F. Haake, and B. Eckhardt, Z. Phys. B {\bf 92},  221  (1993).

\bibitem{Braun:Gerwinski:1996}
P. {Braun}, P. Gerwinski, F. Haake, and H. Schomerus, Z. Phys. B {\bf 100},
  115  (1996).

\bibitem{Pearcey:1946}
T. Pearcey, Philos. Mag. {\bf 37},  311  (1946).

\bibitem{Newton} A.~Mostowski, M.~Stark, {\sl Introduction to higher
algebra}, Pergamon Press, Oxford, 1964.

\bibitem{Bogomolny:Bohigas:1992}
E. Bogomolny, O. Bohigas, and P. Leboeuf, Phys. Rev. Lett. {\bf 68},  2726
  (1992).
\bibitem{HKSSZ}
F. Haake, M. Ku\'s, H.-J. Sommers, H. Schomerus, and K. \.Zyczkowski, J.
Phys. A {\bf 29}, 3641 (1996).

\bibitem{Gerwinski:Haake:1995}
P. Gerwinski, F. Haake, H. Wiedemann, M. Ku\'s, K. \.Zyczkowski,
Phys. Rev. Lett. {\bf 74},  1562  (1995).

\end{references}
\end{document}